\documentclass{article}

\usepackage{arxiv}

\usepackage[utf8]{inputenc} 
\usepackage[T1]{fontenc}    
\usepackage{url}            
\usepackage{booktabs}       
\usepackage{amsfonts}       
\usepackage{nicefrac}       
\usepackage{microtype}      
\usepackage{lipsum}		
\usepackage{graphicx}
\usepackage{doi}

\usepackage{amsmath, amssymb, latexsym}
\usepackage[version=4]{mhchem}
\usepackage{tablefootnote}

\title{Validity of the total quasi-steady-state approximation in stochastic biochemical reaction networks}

\author{ \href{https://orcid.org/0000-0001-7562-5935}{\includegraphics[scale=0.06]{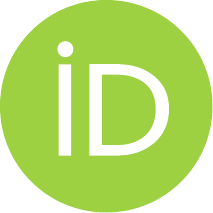}\hspace{1mm}Yun Min Song}\\
	Department of Mathematical Sciences\\ KAIST\\
    Biomedical Mathematics Group\\Pioneer Research Center for\\ Mathematical and Computational Sciences\\
 Institute for Basic Science
	\\
	\texttt{97sym@kaist.ac.kr} \\
	\And
	\href{https://orcid.org/0009-0001-5846-8285}{\includegraphics[scale=0.06]{orcid.pdf}\hspace{1mm}Kangmin Lee} \\
	Department of Mathematical Sciences\\ KAIST\\
    Biomedical Mathematics Group\\Pioneer Research Center for\\ Mathematical and Computational Sciences\\
 Institute for Basic Science\\
	\texttt{lkm0601@kaist.ac.kr} \\
    \And
	\href{https://orcid.org/0000-0001-7842-2172}{\includegraphics[scale=0.06]{orcid.pdf}\hspace{1mm}Jae Kyoung Kim}\thanks{Corresponding author} \\
	Department of Mathematical Sciences\\ KAIST\\
    Biomedical Mathematics Group\\Pioneer Research Center for\\ Mathematical and Computational Sciences\\
 Institute for Basic Science\\
	\texttt{jaekkim@kaist.ac.kr} \\
}

\renewcommand{\shorttitle}{Validity of the stochastic total QSSA}

\hypersetup{
pdftitle={Validity of the total quasi-steady-state approximation in stochastic biochemical reaction networks},
pdfsubject={q-bio.QM},
pdfauthor={Yun Min Song, Kangmin Lee, Jae Kyoung Kim},
pdfkeywords={First keyword, Second keyword, More},
}

\begin{document}
\maketitle

\begin{abstract}
	Stochastic models for biochemical reaction networks are widely used to explore their complex dynamics but face significant challenges, including difficulties in determining rate constants and high computational costs. To address these issues, model reduction approaches based on deterministic quasi-steady-state approximations (QSSA) have been employed, resulting in propensity functions in the form of deterministic non-elementary reaction functions, such as the Michaelis-Menten equation. In particular, the total QSSA (tQSSA), known for its accuracy in deterministic frameworks, has been perceived as universally valid for stochastic model reduction. However, recent studies have challenged this perception. In this review, we demonstrate that applying tQSSA in stochastic model reduction can distort dynamics, even in cases where the deterministic tQSSA is rigorously valid. This highlights the need for caution when using deterministic QSSA in stochastic model reduction to avoid erroneous conclusions from model simulations.
\end{abstract}

\section{Introduction}
Understanding the complex dynamics of biochemical reaction networks, which are fundamental to cellular processes, relies heavily on mathematical modeling \cite{tyson2020dynamical}. For systems with homogeneous spatial distributions of molecules, models based on ordinary differential equations (ODEs) are widely used \cite{mathematicalmodels}. These models represent molecular concentrations as variables and define reaction rates through mass-action kinetics. In contrast, for systems exhibiting spatial heterogeneity, partial differential equations (PDEs) are employed to incorporate spatial variability \cite{mathematicalmodels}.

When molecular copy numbers are too low to support a continuous concentration-based description, stochastic effects become significant \cite{stochasticgeneexpression}. In such cases, stochastic models based on continuous-time Markov chains (CTMCs) provide a more appropriate framework \cite{wilkinson2018stochastic}. These models, like their deterministic counterparts, generally use mass-action kinetics to define propensity functions that govern reaction probabilities. However, analytical solutions for the probability distributions of stochastic models are rarely feasible \cite{schnoerr2017approximation}, necessitating simulation approaches such as the Gillespie algorithm \cite{gillespie1977exact}. For systems with pronounced spatial heterogeneity, compartment-based spatial stochastic simulation algorithms (spatial SSAs) are commonly used to enhance these simulations \cite{erban2007practicalguidestochasticsimulations}.

Stochastic simulations of mass-action kinetics-based models are powerful tools for studying biochemical systems but face significant challenges. Accurately determining reaction rate constants remains difficult \cite{kitano2002computational}, and systems with disparate reaction timescales often require repeated simulations of fast reactions to capture the dynamics of slower processes \cite{gillespie2007stochastic}. To mitigate these challenges, simplification techniques such as the quasi-steady-state approximation (QSSA) are employed \cite{rao2003stochastic, cao2005slow, cao2005accelerated, gomez2008enhanced}. The QSSA reduces computational complexity by neglecting fast reactions and approximating fast-scale variables in propensity functions with their stochastic QSSAs, expressed as their moments conditioned on slow variables. However, stochastic QSSAs are often analytically intractable \cite{schnoerr2017approximation}. Consequently, they are frequently replaced with deterministic QSSAs derived from deterministic models \cite{kim2020misuse}. This substitution results in propensity functions that resemble concentration-dependent non-elementary reaction functions, such as Michaelis-Menten or Hill functions, transformed into count-based forms. This approach has been widely adopted in numerous studies to explore the stochastic dynamics of biochemical reaction networks. \cite{gonze2002biochemical, gonze2002deterministic, pedraza2005noise, scott2006estimations, tian2006stochastic, ccaugatay2009architecture, ouattara2010structure, gonze2011molecular, smadbeck2012stochastic, riba2014combination, dovzhenok2015mathematical, zhang2015negative, schuh2020gene, jia2020small, bersani2008deterministic, choi2017beyond, d2017stability, beesley2020wake, bersani2020stochastic, barik2008stochastic, sanft2011legitimacy, kang2019quasi}.

For stochastic model reduction using deterministic QSSA, various forms of deterministic QSSA, including the standard QSSA (sQSSA) and total QSSA (tQSSA), are available \cite{kim2014validity, kim2015relationship, kim2020misuse}. Among these, the tQSSA has demonstrated superior performance in accurately capturing the original deterministic dynamics \cite{kim2020misuse}. For example, while the sQSSA, which leads to the Michaelis-Menten (MM) equations, often distorts the dynamics of the original deterministic system, the tQSSA has proven to be more accurate across a broader range of conditions in deterministic ODE models \cite{kim2020misuse} and has improved the fidelity of PDE model simplifications \cite{BeyondMMPDE}. This enhanced accuracy in deterministic models likely explains the superior performance of stochastic simulations employing tQSSA-based equations (stochastic tQSSA) compared to those using other deterministic QSSAs \cite{kim2014validity, kim2015relationship, kim2020misuse, barik2008stochastic}. As a result, the stochastic tQSSA has been widely regarded as reliable \cite{kim2014validity,kim2015relationship, kim2020misuse,barik2008stochastic,macnamara,JITHINRAJ20141,Herath,cao2005accelerated, sanft2011legitimacy, kang2019quasi, kim2017reductionconserv}, particularly in scenarios where the deterministic tQSSA is rigorously valid \cite{noethen2011tikhonov}, such as in rapid reversible binding processes.

However, recent studies challenge this assumption, revealing that the stochastic tQSSA is not universally valid \cite{kim2017reduction, song2021Universally, Chae2025homohete}. This review critically examines these limitations, showing that stochastic tQSSA can introduce significant distortions in the dynamics of simple gene regulatory network models, even in cases where the deterministic tQSSA accurately captures the original dynamics. These distortions are evident in both homogeneous and heterogeneous spatial contexts, emphasizing the need for caution when applying tQSSA in stochastic simulations.

\section{Results}
\label{sec:headings}

\subsection{Deterministic and stochastic tQSSA for gene regulatory network dynamics under spatial homogeneity}

To evaluate the validity of the stochastic tQSSA, we first derive the deterministic tQSSA and demonstrate its application in simplifying the stochastic model for a gene regulation system. In this model, mRNA (M) is transcribed from DNA (D) at a rate \(\alpha_m\) and degraded at a rate \(d_m\). D reversibly binds to a repressor (P) to form a complex (D:P), which inhibits mRNA transcription. The binding and unbinding rates are denoted by \(k_f\) and \(k_b\), respectively. Assuming homogeneous spatial conditions, the deterministic dynamics of the species' concentrations can be expressed using mass-action-based ordinary differential equations (ODEs):  

\begin{equation}
\begin{aligned}
\frac{dD}{dt} &= -k_f D \cdot P + k_b D\text{:}P,\\
\frac{dP}{dt} &= -k_f D \cdot P + k_b D\text{:}P,\\
\frac{dD\text{:}P}{dt} &= k_f D \cdot P - k_b D\text{:}P,\\
\frac{dM}{dt} &= \alpha_m D - d_m M. \label{eq:odefull}
\end{aligned}
\end{equation}
By introducing the variables representing the total DNA ($D_T= D+D\text{:}P$) and repressor ($P_T= P+D\text{:}P$) concentrations instead of $P$ and $D\text{:}P$, we can rewrite the ODE system as follows:
\begin{equation}
\begin{aligned}
\frac{dD}{dt} &= -k_f D \cdot (P_T - D_T + D) + k_b (D_T - D),\\
\frac{dM}{dt} &= \alpha_m D - d_m M,\\
\frac{dD_T}{dt} &= 0,\\
\frac{dP_T}{dt} &= 0.
\end{aligned} \label{eq:odefull2}
\end{equation}
Here the time derivatives of $D_T$ and $P_T$ are zero, indicating that these quantities remain constant over time.

Generally, the reactions involving the binding and unbinding between DNA and repressor occur in much faster time scales than mRNA production (i.e., $k_f, k_b \gg \alpha_m, d_m$). Then $D$, regulated by only the fast reversible binding reactions, evolves in a much faster time scale than the other slow (or fixed) variables ($M$, $D_T$, and $P_T$), not regulated by the fast reactions. Then, in the fast time scale, $D$ quickly converges to its quasi-steady-state (QSS) while the other slow variables ($M$, $D_T$, and $P_T$) remain constant. The approximation of the QSS (QSSA; $D_{tq}$) can be obtained by solving $dD/dt = 0$:
\begin{equation}
\begin{aligned}
D_{tq}(D_T, P_T) = \frac{1}{2} \left \{(D_T - P_T - K_d) + \sqrt{(D_T - P_T - K_d)^2 + 4D_T K_d} \right \},
\end{aligned}\label{eq:tQSSA}
\end{equation}
where $K_d = k_b/k_f$ is the dissociation constant. The derived QSSA is called the total QSSA (tQSSA) as it is a function in terms of the total variables (i.e., $D_T$, and $P_T$). Then we can derive the reduced model by replacing $D$ by the tQSSA ($D_{tq}$)
\begin{equation}
\begin{aligned}
\frac{dM}{dt} &= \alpha_m D_{tq}(D_T, P_T) - d_m M,\\
\frac{dD_T}{dt} &= 0,\\
\frac{dP_T}{dt} &= 0. \label{eq:odereduced}
\end{aligned}
\end{equation}
This reduction is mathematically justified by Tikhonov's theorem \cite{noethen2011tikhonov}. 


When the molecule copy numbers of the species are low, the stochastic dynamics of their interactions, which cannot be described by deterministic models, can no longer be ignored. To model these stochastic dynamics, the system can be described by a CTMC of the molecule copy numbers \( n_X = X \cdot \Omega \), where \( X \) represents the species \( D \), \( P \), \( D\text{:}P \), or \( M \), and \( \Omega \) is the system volume. The propensity functions for the reactions that alter the copy numbers follow a mass-action form (Table~\ref{table:1}).

\begin{table}[h]
\begin{center}
\caption{Propensity functions of the full gene regulation model under spatial homogeneity.}
\begin{tabular}{|c|c|} \hline
Reactions & Propensities \\ \hline \hline
\ce{D + P -> D\text{:}P} & $\frac{k_f}{\Omega} n_D\cdot n_P$  \\ \hline
\ce{D\text{:}P -> D + P} & $k_b n_{D\text{:}P}$  \\ \hline
\ce{D -> D + M} & $\alpha_m n_D$  \\ \hline
\ce{M -> $\emptyset$} & $d_m n_M$ \\ \hline
\end{tabular}
\label{table:1}
\end{center}
\end{table}

Simulating this model using the Gillespie algorithm allows analysis of stochastic dynamics. However, the computational cost of such stochastic simulations is high, hampering the analysis of dynamics and parameter estimation that require numerous simulations. Thus, simplifying the stochastic model is highly favorable. To simplify the stochastic model, one can consider the stochastic counterpart of the deterministic reduced model (Eq.~\eqref{eq:odereduced}), where all fast reactions are eliminated and the fast variable \( n_D \) in the propensity functions is substituted by the tQSSA equation (Table~\ref{table:2}). In this process, all concentration-dimension variables are scaled by \( \Omega \) to convert them into count-dimension variables:

\begin{table}[h]
\begin{center}
\caption{Propensity functions of the reduced gene regulation model under spatial homogeneity.}
\begin{tabular}{|c|c|} \hline
Reactions & Propensities \\ \hline \hline
\ce{ $\emptyset$ -> M} & $ \alpha_m n_{D_{tq}}(n_{D_T}, n_{P_T})$ \\ \hline
\ce{M -> $\emptyset$} & $d_m n_M$  \\ \hline 
\end{tabular}
\label{table:2}
\end{center}
\end{table}
Here,
\begin{equation}
\begin{aligned}
n_{D_{tq}}(n_{D_T}, n_{P_T}) &= D_{tq} \cdot \Omega\\
&= \frac{1}{2} \left \{(n_{D_T} - n_{P_T} - K_d\cdot\Omega) + \sqrt{(n_{D_T} - n_{P_T} - K_d\cdot\Omega)^2 + 4n_{D_T} K_d\Omega} \right \}, \label{eq:stQSSA}
\end{aligned}
\end{equation}
where $n_{D_T} = n_D + n_{D\text{:}P}$ and $n_{P_T} = n_P + n_{D\text{:}P}$. The derived equation approximating the stochastic QSS (Eq.~\eqref{eq:stQSSA}) is commonly referred to as the stochastic tQSSA. By eliminating the fast reactions from the full stochastic model (Table~\ref{table:1}; the first two reactions), the reduced model (Table~\ref{table:2}) requires significantly fewer reactions for simulation over the same time frame. As a result, the reduced model achieves substantially lower computational costs compared to the full model.

\subsection{Stochastic tQSSA can distort gene regulation network dynamics under spatial homogeneity}

\begin{figure}[ht]
 \begin{center}
 \includegraphics[scale = 0.7]{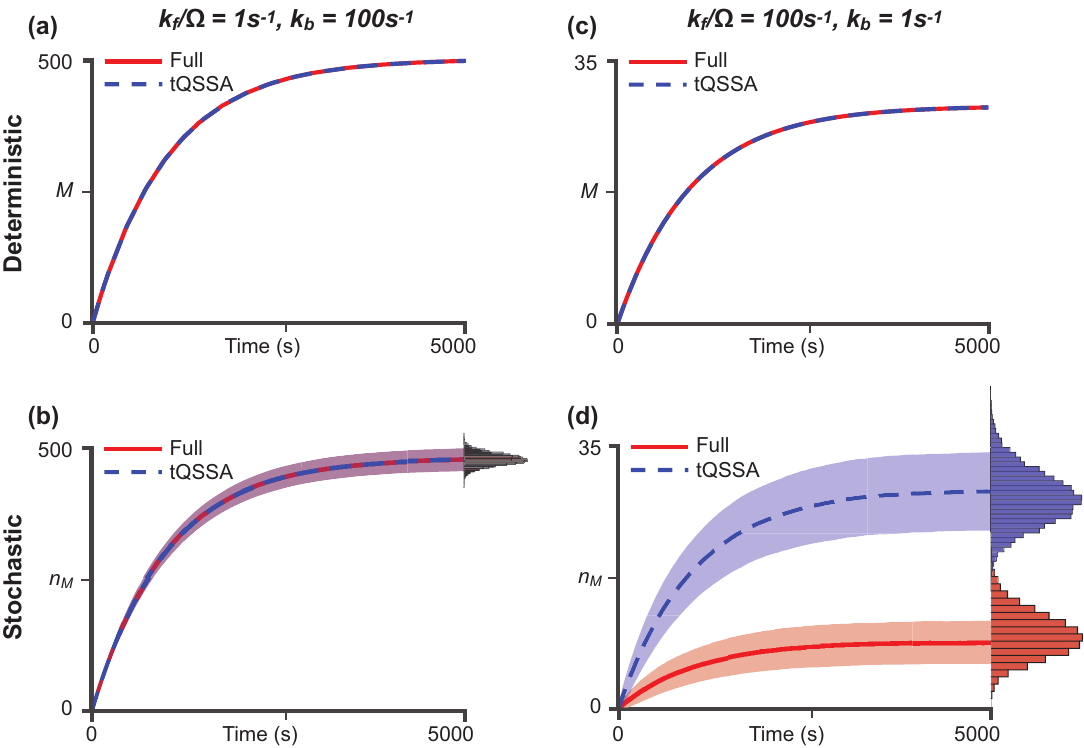}
 \caption{ Stochastic tQSSA can distort dynamics even when the deterministic tQSSA is valid under spatial homogeneity. \textbf{(a)} When the binding (\(k_f / \Omega = 1~\mathrm{s^{-1}}\)) and unbinding (\(k_b = 100~\mathrm{s^{-1}}\)) rates are much faster than the other reactions (\(\alpha_m = 0.1~\mathrm{s^{-1}}\) and \(d_m = 0.001~\mathrm{s^{-1}}\)), \(M\) simulated with the deterministic full model (Eq.~\ref{eq:odefull}, red solid line) and the reduced model (Eq.~\ref{eq:odereduced}, blue dashed line) precisely match. \textbf{(b)} Under the same conditions, \(n_M\) simulated with the stochastic full model (Table~\ref{table:1}, red solid line) and the reduced model (Table~\ref{table:2}, blue dashed line) also match closely, as various prior studies expected. Here, the lines with shaded regions represent the mean \(\pm\) standard deviation, and the histograms depict the stationary distribution of \(10^4\) trajectories. \textbf{(c-d)} However, when the amounts of the rapidly binding species are similar (\(n_{D_T} = n_{P_T} = 10\)) and the binding becomes tight (\(k_f / \Omega = 100~\mathrm{s^{-1}}, k_b = 1~\mathrm{s^{-1}}\)), the deterministic full and reduced models still precisely match (c), but the stochastic reduced model fails to replicate the dynamics of the full model (d). Here, \(\Omega = 1\) (arbitrary unit), and the initial condition is \([D, P, D\text{:}P, M] = [10, 10, 0, 0]\).
}
 \label{fig:1}
 \end{center}
\end{figure}

Under the assumption of rapid reversible binding (i.e., \(k_f, k_b \gg \alpha_m, d_m\)), the deterministic tQSSA (Eq.~\eqref{eq:tQSSA}) accurately captures the deterministic QSS of \(D\) \cite{noethen2011tikhonov, borghans1996extending, schnell2002enzyme, tzafriri2003michaelis}. Consequently, when simulating the deterministic full (Eq.~\eqref{eq:odefull}) and reduced (Eq.~\eqref{eq:odereduced}) models under conditions of binding and unbinding rates much higher than the other reaction rates, the reduced model accurately replicates the slow dynamics of the full model (Fig.~\ref{fig:1}a). Such valid reduction by using the deterministic tQSSA has led to expectations in previous studies that the reduced model by using the stochastic tQSSA (Table~\ref{table:2}) would similarly replicate the slow dynamics of the stochastic full model (Table~\ref{table:1}) \cite{kim2014validity,kim2015relationship, kim2020misuse,barik2008stochastic,macnamara,JITHINRAJ20141,Herath,cao2005accelerated, sanft2011legitimacy, kang2019quasi, kim2017reductionconserv}. Indeed, for most cases where the deterministic tQSSA is valid, the stochastic reduced model effectively captures the slow dynamics of the stochastic full model (Fig.~\ref{fig:1}b).

However, a recent study revealed that the stochastic tQSSA equation (Eq.~\eqref{eq:stQSSA}) substantially overestimates the stochastic QSS under specific conditions: when \(n_{D_T} K_d \Omega < 10\) and \(n_{D_T} \approx n_{P_T}\), even with rapid reversible binding \cite{song2021Universally}. The first condition arises in scenarios of tight molecular binding (i.e., small \(K_d\)) combined with low molecular copy numbers (\(n_{D_T}\)) and small system volumes (\(\Omega\)), where stochastic effects dominate and the system deviates from deterministic behavior. The second condition occurs when the binding species have comparable copy numbers. Notably, this invalid condition for the stochastic tQSSA is independent of the deterministic tQSSA’s validity. Thus, under the condition, the deterministic tQSSA remains valid (Fig.~\ref{fig:1}c). In contrast, the stochastic reduced model exhibits significant discrepancies compared to the stochastic full model (Fig.~\ref{fig:1}d). Specifically, the stochastic tQSSA equation (Eq.~\eqref{eq:stQSSA})  overestimates the stochastic QSS, resulting in an inflated production rate and a distorted stationary distribution of M.

These results highlight that while tQSSA-based reductions are effective in many scenarios, they are not universally applicable. Furthermore, the validity conditions for applying the tQSSA in stochastic systems are stricter than those for deterministic systems, requiring careful consideration in stochastic modeling. In such cases, an alternative QSSA, introduced in a previous study \cite{song2021Universally} and referred to as the stochastic low-state QSSA (lQSSA), offers a valid approach for stochastic model reduction.

\subsection{Deterministic and stochastic tQSSA for gene regulatory network dynamics under spatial heterogeneity}
Next, we derive the deterministic tQSSA and demonstrate its application in simplifying the stochastic model in a spatially heterogeneous context. Building on the gene regulation model from the previous section, we introduce diffusion for P and M in a bounded one-dimensional domain. D and D:P are assumed not to diffuse as the DNA remains localized within the nucleus. This system can be described by partial differential equations (PDEs) based on mass-action kinetics:

\begin{equation}
\begin{aligned}
\frac{\partial D}{\partial t} &= -k_f D \cdot P + k_b D\text{:}P,\\
\frac{\partial P}{\partial t} &= \delta_P \Delta P -k_f D \cdot P + k_b D\text{:}P,\\
\frac{\partial D\text{:}P}{\partial t} &= k_f D \cdot P - k_b D\text{:}P,\\
\frac{\partial M}{\partial t} &= \delta_M\Delta M + \alpha_m D - d_m M, \label{eq:pdefull}
\end{aligned}
\end{equation}
where $\delta_X$ denotes the diffusion coefficient of the corresponding species ($X=P, M$). Similar to the corresponding ODE model (Eq.~\ref{eq:odefull}), the full model can be rewritten by introducing $D_T = D + D\text{:}P$ and $P_T = P + D\text{:}P$:
\begin{equation}
\begin{aligned}
\frac{\partial D}{\partial t} &= -k_f D \cdot (P_T - D_T + D) + k_b (D_T - D),\\
\frac{\partial M}{\partial t} &= \delta_M\Delta M + \alpha_m D - d_m M,\\
\frac{\partial P_T}{\partial t} &= \delta_P \Delta (P_T-D_T + D),\\
\frac{\partial D_T}{\partial t} &= 0. \label{eq:pdefull2}
\end{aligned}
\end{equation}
Notably, unlike the corresponding ODE model (Eq.~\ref{eq:odefull}), the $P_T$ is time-variant. 

If the diffusion coefficients are large or comparable to the reaction rate constants of the fast reversible binding reactions (i.e., $k_f$ and $k_b$), the spatial heterogeneity is quickly resolved, so that the PDE model (Eq.~\ref{eq:pdefull}) becomes nearly equivalent to the corresponding ODE model (Eq.~\ref{eq:odefull}). Thus, we assume that diffusion coefficients are comparable or much smaller than the slow reaction rate constants (i.e., $\alpha_m$ and $d_m$). Under this condition, we can assume that $D$ quickly converges to the QSS while the other variables remain constants, where the QSSA of $D$ is equivalent to that obtained from the ODE model (i.e., $D_{tq}$ (Eq.~\ref{eq:tQSSA})). Thus, the full PDE model can be simplified using the tQSSA \cite{frank2018quasi,kalachev2007reduction,Chae2025homohete}:
\begin{equation}
\begin{aligned}
\frac{\partial M}{\partial t} &= \delta_M\Delta M + \alpha_m D_{tq} - d_m M,\\
\frac{\partial P_T}{\partial t} &= \delta_P \Delta (P_T-D_T + D_{tq}),\\
\frac{\partial D_T}{\partial t} &= 0.
\end{aligned} \label{eq:pdereduced}
\end{equation}

For simulations involving low molecular copy numbers, a compartment-based Gillespie algorithm can be employed \cite{erban2007practicalguidestochasticsimulations, Chae2025homohete}. In this approach, the one-dimensional spatial domain is discretized into \( n \) compartments, with reactions occurring independently within each compartment (Table~\ref{table:3}). The reaction propensities in the \( i \)-th compartment are determined by the copy numbers of the reactants within that compartment (\(n_{X,i}; X = D, P, D\text{:}P, M\)). Assuming the length of the bounded domain is \( L \), the length of each compartment is \( h=L/n \). Additionally, species diffusion between adjacent compartments is modeled as inter-compartmental conversion reactions. For instance, the diffusion of P from the \( i \)-th compartment to the (\( i+1 \))-th compartment is represented by the conversion reaction \( \ce{P_i -> P_{i+1}} \). At the boundary compartments (the first and \( n \)-th compartments), diffusion is restricted to the adjacent compartments (the second and (\( n-1 \))-th compartments, respectively). Since the length of each compartment is \( h \), the conversion reaction rate constant across compartments is \( \tilde{\delta}_X = \delta_X/h^2 \), where \( X = P, M \) \cite{erban2007practicalguidestochasticsimulations, Chae2025homohete}.

\begin{table}[h]
\begin{center}
\caption{Propensity functions of the full gene regulation model under spatial heterogeneity.}
\begin{tabular}{|c|c|} \hline
Reactions & Propensities \\ \hline \hline
\ce{D_i + P_i -> D\text{:}P_i}, $i=1,...,n$ & $\frac{k_f}{\Omega_i} n_{D,i}\cdot n_{P,i}$  \\ \hline
\ce{D\text{:}P_i -> D_i + P_i}, $i=1,...,n$ & $k_b n_{D\text{:}P, i}$  \\ \hline
\ce{D_i -> D_i + M_i}, $i=1,...,n$ & $\alpha_m n_{D,i}$  \\ \hline
\ce{M_i -> $\emptyset$}, $i=1,...,n$ & $d_m n_{M,i}$ \\ \hline
\ce{P_i -> P_{i+1}}, $i=1,...,(n-1)$ & $\tilde{\delta}_P n_{P,i}$ \\ \hline
\ce{P_i -> P_{i-1}}, $i=2,...,n$ & $\tilde{\delta}_P n_{P,i}$ \\ \hline
\ce{M_i -> M_{i+1}}, $i=1,...,(n-1)$ & $\tilde{\delta}_M n_{M,i}$ \\ \hline
\ce{M_i -> M_{i-1}}, $i=2,...,n$ & $\tilde{\delta}_M n_{M,i}$ \\ \hline
\end{tabular}
\label{table:3}
\end{center}
\end{table}

Similar to the spatially homogeneous case, we can consider the stochastic counterpart of the reduced PDE model as a reduced model of the full spatial stochastic system \cite{Chae2025homohete} (Table~\ref{table:4}). In this reduced model, all fast reactions are removed, and the fast variable \( n_{D,i} \) in the propensity functions is replaced by the stochastic tQSSA for each compartment:
\begin{equation}
\begin{aligned}
&n_{D_{tq},i}(n_{D_T,i}, n_{P_T,i}) \\
&= \frac{1}{2} \left \{(n_{D_T,i} - n_{P_T,i} - K_d\cdot\Omega_i) + \sqrt{(n_{D_T,i} - n_{P_T,i} - K_d\cdot\Omega_i)^2 + 4n_{D_T} K_d\Omega_i} \right \}, \label{eq:stQSSA_i}
\end{aligned}
\end{equation}
where $n_{D_T,i} = n_{D,i} + n_{D\text{:}P,i}$, $n_{P_T,i} = n_{P,i} + n_{D\text{:}P,i}$, and $\Omega_i$ is the compartment volume.
Simulating this reduced model is significantly more efficient than the full model due to the elimination of all fast reactions, as in the spatially homogeneous case.

\begin{table}[h]
\begin{center}
\caption{Propensity functions of the reduced gene regulation model under spatial heterogeneity.}
\begin{tabular}{|c|c|} \hline
Reactions & Propensities \\ \hline \hline
\ce{ $\emptyset$ -> M_i} & $\alpha_M n_{D_{tq}}(n_{D_T, i}, n_{P_T, i})$   \\ \hline
\ce{M_i -> $\emptyset$} & $d_m n_{M,i}$  \\ \hline 
\ce{P_{T}_{i} -> P_{T}_{i+1}}, $i=1,...,(n-1)$& 
$\tilde{\delta}_P (n_{P_{T},i}-n_{D_{tq}}(n_{D_T, i}, n_{P_T, i})$ \\ \hline
\ce{P_{T}_{i} -> P_{T}_{i-1}}, $i=2,...,n$ & $\tilde{\delta}_P (n_{P_T,i}-n_{D_{tq}}(n_{D_T, i}, n_{P_T, i})$ \\ \hline
\ce{M_i -> M_{i+1}}, $i=1,...,(n-1)$ & $\tilde{\delta}_M n_{M,i}$ \\ \hline
\ce{M_i -> M_{i-1}}, $i=2,...,n$ & $\tilde{\delta}_M n_{M,i}$ \\ \hline
\end{tabular}
\label{table:4}
\end{center}
\end{table}

\subsection{Stochastic tQSSA can distort gene regulation network dynamics under spacial heterogeneity}

\begin{figure}[ht]
 \begin{center}
 \includegraphics[scale = 0.75]{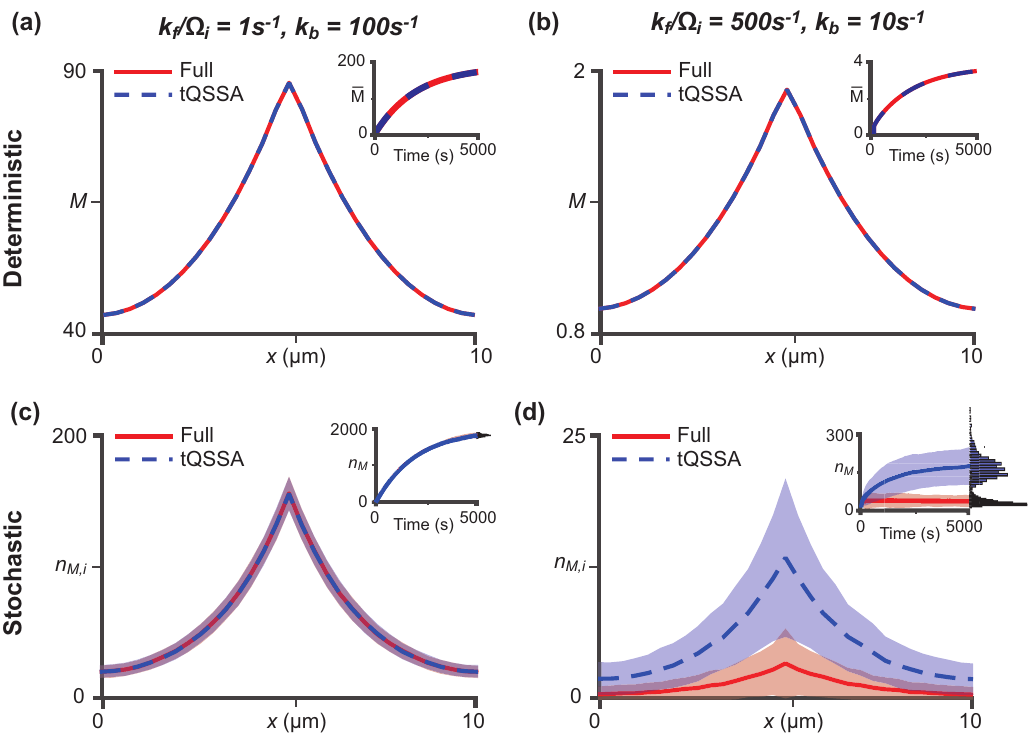}
 \caption{Stochastic tQSSA can distort dynamics even when deterministic tQSSA is valid under spatial heterogeneity.  
\textbf{(a)} When the binding (\(k_f / \Omega_i = 1~\mathrm{s^{-1}}\)) and unbinding (\(k_b = 100~\mathrm{s^{-1}}\)) rates are much faster than the other reactions (\(\alpha_m = 0.5~\mathrm{s^{-1}}, d_m = 0.0005~\mathrm{s^{-1}}\)) and diffusion (\(\delta_M = \delta_P = 0.002~\mu\mathrm{m^2/s}\)), \(M\) at \(t=5000\) and the spatial mean \(M\) trajectory ($\overline{M}$; inset) simulated with the deterministic full model (Eq.~\ref{eq:pdefull}, red solid line) and the reduced model (Eq.~\ref{eq:pdereduced}, blue dashed line) are in precise alignment.  
\textbf{(b)} Under the same conditions, \(n_{M,i}\) at \(t=5000\) and the spatial total \(n_{M,i}\) trajectory ($n_M$; inset) simulated with the stochastic full model (Table~\ref{table:3}, red solid line) and the reduced model (Table~\ref{table:4}, blue dashed line) also show close agreement. Here, $n_{M,i}$ were plotted at the center of the corresponding compartment on the $x$-axis and then interpolated. The lines with shaded regions represent the mean \(\pm\) standard deviation, and the histograms illustrate the stationary distribution from \(10^3\) trajectories.  
\textbf{(c-d)} However, when binding becomes tight (\(k_f / \Omega_i = 500~\mathrm{s^{-1}}, k_b = 10~\mathrm{s^{-1}}\)), the deterministic full and reduced models still match precisely (c), but the stochastic reduced model fails to replicate the dynamics of the full model (d). Notably, while the spatial total amounts of the reversibly binding species are not comparable (\(n_{P_T} = 31\) and \(n_{D_T} = 2\)), their local amounts become comparable (\(n_{P_T,16} \approx 1\) and \(n_{D_T,16} = 2\)), causing local violations of the stochastic tQSSA.  
Here, \(L = 10~\mu\mathrm{m}\), \(n = 31\), and $\Omega_i = 1$ (arbitrary unit). The initial conditions are \(D(x,0) = 2I_{\{5-h/2 < x < 5+h/2\}}(x)\), \(P(x,0) = I_{\{0 < x < 10\}}(x)\), and \(D{:}P(x,0) = M(x,0) = 0\).
 }
 \label{fig:2}
 \end{center}
\end{figure}

When the rapid reversible binding is much faster than the other reactions and diffusion ($k_f, k_b \gg \alpha_m, d_m, \delta_M, \delta_P$), the time scale of $D$ and the other variables are well separated in the full PDE model (Eq.~\eqref{eq:pdefull2}). Thus, the deterministic tQSSA (Eq.~\eqref{eq:tQSSA}) is expected to accurately captures the deterministic QSS of \(D\), even under spacial heterogeneity \cite{frank2018quasi,kalachev2007reduction, Chae2025homohete}. Indeed, when simulating the deterministic full (Eq.~\eqref{eq:pdefull}) and reduced models (Eq.~\eqref{eq:pdereduced}) with binding and unbinding rates much faster than other reactions and diffusion, the reduced model accurately reproduces the dynamics of \(M\) in the full model (Fig.~\ref{fig:2}a). In this simulation, the models assumed a one-dimensional spatial domain of length \(L = 10~\mu m\), with D localized in the central \(10/31~\mu m\) region, and Neumann boundary conditions. The valid reductions using the deterministic tQSSA in such cases has led to the expectation that it would also hold in the stochastic context. Indeed, for various scenarios meeting the deterministic tQSSA validity conditions, the stochastic reduced model (Table~\ref{table:4}) accurately captures the slow dynamics of the full model (Table~\ref{table:3}) (Fig.~\ref{fig:2}b). In this simulation, the number of compartments was 31 and the D was localized only in the single center compartment. 

However, as shown in a previous section, the stochastic tQSSA is not universally valid, even under conditions of rapid reversible binding \cite{song2021Universally}. Moreover, unlike the spatially homogeneous case, the validity of the stochastic tQSSA in spatially heterogeneous systems must be evaluated locally \cite{Chae2025homohete}. Specifically, while the total amounts of the two binding species across the entire spatial domain (i.e., \(n_{D_T}\) and \(n_{P_T}\) where \(n_X = \sum_{i=1}^{n} n_{X,i}\)) may differ significantly, their local amount can be comparable in certain compartments (i.e., \(n_{D_T,i}\) and \(n_{P_T,i}\)). Furthermore, these compartments typically have smaller volumes (\(\Omega_i\)) and lower copy numbers (\(n_{D_T,i}\)) compared to the whole system. Consequently, the invalid condition of the stochastic tQSSA may be satisfied locally (i.e., $n_{D_T,i} K_d\Omega_i < 10$ and $n_{D_T,i} \approx n_{P_T,i}$), even if it is not met in the corresponding spatially homogeneous system. Indeed, when the rapid reversible binding becomes tight, the deterministic reduced model remains valid (Fig.~\ref{fig:2}c), but the stochastic reduced model fails to capture the dynamics of the original system (Fig.~\ref{fig:2}d). Specifically, even if the total counts of D and P across the spatial domain differ significantly in the system ($n_{D_T} = 2, n_{P_T} = 31$), one of the 31 compartments, where D is localized, experiences comparable species counts ($n_{D_T,i} = 2, n_{P_T,i} \approx 1$). This local violation of the stochastic tQSSA validity condition leads to an overestimation of the production rate of M, ultimately distorting the original system's dynamics.

These results emphasize that, similar to the spatially homogeneous case, tQSSA-based stochastic model reductions are not universally applicable, with stricter validity conditions than in deterministic models. Moreover, caution is particularly necessary under spatial heterogeneity, as it can lead to local violations of validity conditions that are otherwise satisfied in spatially homogeneous systems. For such cases, especially in compartments where the stochastic tQSSA validity condition is locally violated, the usage of the alternative QSSA (lQSSA) is required \cite{song2021Universally, Chae2025homohete}.

\section{Discussion}

In this review, we examined the limitations of the stochastic tQSSA, which has been widely regarded as an efficient and accurate tool for stochastic simulations \cite{gonze2002biochemical, gonze2002deterministic, pedraza2005noise, scott2006estimations, tian2006stochastic, ccaugatay2009architecture, ouattara2010structure, gonze2011molecular, smadbeck2012stochastic, riba2014combination, dovzhenok2015mathematical, zhang2015negative, schuh2020gene, jia2020small, bersani2008deterministic, choi2017beyond, d2017stability, beesley2020wake, bersani2020stochastic, barik2008stochastic, sanft2011legitimacy, kang2019quasi}. Specifically, we demonstrated that applying the stochastic tQSSA can significantly distort dynamics, even when the deterministic tQSSA accurately captures the original system’s behavior (i.e., fast binding and unbinding reactions). This underscores that the stochastic tQSSA has stricter validity conditions than its deterministic counterpart, warranting cautious application.

The invalidity condition of the stochastic tQSSA, identified in a previous study \cite{song2021Universally} as tight binding between species with comparable counts, seems relatively narrow. This may explain why earlier studies often observed good agreement between the stochastic tQSSA and the original dynamics. However, the tight binding condition (\(n_{D_T}K_d \Omega < 10\)), which defines this invalidity, is plausible in real biological systems \cite{song2021Universally}. Moreover, as systems evolve, they may transiently enter invalid regimes, characterized by comparable amounts of reversibly binding species \cite{song2021Universally}. Additionally, spatial heterogeneity can locally violate the validity condition even when the corresponding spatially homogeneous system remains valid, as demonstrated in our Results section and previous work \cite{Chae2025homohete}.
Therefore, although the invalidity condition appears narrow, it requires cautious application.

In such scenarios, the stochastic low-state QSSA (lQSSA), proposed in earlier work \cite{song2021Universally}, serves as a viable alternative. The stochastic lQSSA is valid under conditions of tight binding (\(n_{D_T}K_d \Omega < 10\)) and can thus address scenarios where the stochastic tQSSA becomes invalid. Consequently, the adaptive application of the stochastic tQSSA and lQSSA enables a universally valid reduction of stochastic models under rapid reversible binding conditions \cite{song2021Universally}. Importantly, in spatially heterogeneous systems, the stochastic tQSSA and lQSSA should be applied on a compartmental basis, determined by local validity criteria (i.e., \(n_{D_T,i}K_d \Omega_i < 10\)).

If the time scales of the variables in a system are not fully separated (e.g., reversible binding is not significantly faster than other reactions), even the deterministic tQSSA may fail to accurately capture the system's dynamics \cite{ETS}. To address such situations, recent studies have developed an effective time-delay scheme (ETS) \cite{ETS} and its extensions \cite{Chae_2025}. This scheme rigorously estimates the time-delay effects in molecular complex formation during reversible binding, which are negligible when the QSSA approach is valid. These studies demonstrated that ETS can accurately replicate the original deterministic dynamics even in cases where tQSSA is invalid. They also applied ETS to a simple stochastic system, assuming only one DNA binding site. Future work could explore whether this scheme remains applicable under spatial heterogeneity or in stochastic models where interacting species with copy numbers greater than one undergo reversible binding. There may be scenarios where ETS fails due to fundamental differences between deterministic and stochastic models, which would warrant further investigation.

\section*{Acknowledgments}
This study was funded by the Institutetute for Basic Science (IBS-R029-C3) (J.K.K.).

\bibliographystyle{unsrt}
\bibliography{manuscript}

\begin{thebibliography}{10}

\bibitem{tyson2020dynamical}
John~J Tyson and Bela Novak.
\newblock A dynamical paradigm for molecular cell biology.
\newblock {\em Trends Cell Biol}, 2020.

\bibitem{mathematicalmodels}
Leah Edelstein-Keshet.
\newblock {\em Mathematical Models in Biology}.
\newblock Society for Industrial and Applied Mathematics, 2005.

\bibitem{stochasticgeneexpression}
Michael~B. Elowitz, Arnold~J. Levine, Eric~D. Siggia, and Peter~S. Swain.
\newblock Stochastic gene expression in a single cell.
\newblock {\em Science}, 297(5584):1183--1186, 2002.

\bibitem{wilkinson2018stochastic}
Darren~J Wilkinson.
\newblock {\em Stochastic modelling for systems biology}.
\newblock Chapman and Hall/CRC, 2018.

\bibitem{schnoerr2017approximation}
David Schnoerr, Guido Sanguinetti, and Ramon Grima.
\newblock Approximation and inference methods for stochastic biochemical kinetics—a tutorial review.
\newblock {\em Journal of Physics A: Mathematical and Theoretical}, 50(9):093001, 2017.

\bibitem{gillespie1977exact}
Daniel~T Gillespie.
\newblock Exact stochastic simulation of coupled chemical reactions.
\newblock {\em The Journal of Chemical Physics}, 81(25):2340--2361, 1977.

\bibitem{erban2007practicalguidestochasticsimulations}
Radek Erban, Jonathan Chapman, and Philip Maini.
\newblock A practical guide to stochastic simulations of reaction-diffusion processes, 2007.

\bibitem{kitano2002computational}
Hiroaki Kitano.
\newblock Computational systems biology.
\newblock {\em Nature}, 420(6912):206--210, 2002.

\bibitem{gillespie2007stochastic}
Daniel~T Gillespie.
\newblock Stochastic simulation of chemical kinetics.
\newblock {\em Annual Review of Physical Chemistry}, 58(1):35--55, 2007.

\bibitem{rao2003stochastic}
Christopher~V Rao and Adam~P Arkin.
\newblock Stochastic chemical kinetics and the quasi-steady-state assumption: Application to the gillespie algorithm.
\newblock {\em The Journal of Chemical Physics}, 118(11):4999--5010, 2003.

\bibitem{cao2005slow}
Yang Cao, Daniel~T Gillespie, and Linda~R Petzold.
\newblock The slow-scale stochastic simulation algorithm.
\newblock {\em The Journal of Chemical Physics}, 122(1):014116, 2005.

\bibitem{cao2005accelerated}
Yang Cao, Daniel~T Gillespie, and Linda~R Petzold.
\newblock Accelerated stochastic simulation of the stiff enzyme-substrate reaction.
\newblock {\em The Journal of Chemical Physics}, 123(14):144917, 2005.

\bibitem{gomez2008enhanced}
Carlos~A G{\'o}mez-Uribe, George~C Verghese, and Abraham~R Tzafriri.
\newblock Enhanced identification and exploitation of time scales for model reduction in stochastic chemical kinetics.
\newblock {\em The Journal of Chemical Physics}, 129(24):244112, 2008.

\bibitem{kim2020misuse}
Jae~Kyoung Kim and John~J Tyson.
\newblock Misuse of the michaelis--menten rate law for protein interaction networks and its remedy.
\newblock {\em PLOS Computational Biology}, 16(10):e1008258, 2020.

\bibitem{gonze2002biochemical}
Didier Gonze, Jos{\'e} Halloy, and Pierre Gaspard.
\newblock Biochemical clocks and molecular noise: Theoretical study of robustness factors.
\newblock {\em The Journal of Chemical Physics}, 116(24):10997--11010, 2002.

\bibitem{gonze2002deterministic}
Didier Gonze, Jos{\'e} Halloy, and Albert Goldbeter.
\newblock Deterministic versus stochastic models for circadian rhythms.
\newblock {\em Journal of Biological Physics}, 28(4):637--653, 2002.

\bibitem{pedraza2005noise}
Juan~M Pedraza and Alexander van Oudenaarden.
\newblock Noise propagation in gene networks.
\newblock {\em Science}, 307(5717):1965--1969, 2005.

\bibitem{scott2006estimations}
Matthew Scott, Brian Ingalls, and Mads K{\ae}rn.
\newblock Estimations of intrinsic and extrinsic noise in models of nonlinear genetic networks.
\newblock {\em Chaos}, 16(2):026107, 2006.

\bibitem{tian2006stochastic}
Tianhai Tian and Kevin Burrage.
\newblock Stochastic models for regulatory networks of the genetic toggle switch.
\newblock {\em Proceedings of the National Academy of Sciences of the United States of America}, 103(22):8372--8377, 2006.

\bibitem{ccaugatay2009architecture}
Tolga {\c{C}}a{\u{g}}atay, Marc Turcotte, Michael~B Elowitz, Jordi Garcia-Ojalvo, and G{\"u}rol~M S{\"u}el.
\newblock Architecture-dependent noise discriminates functionally analogous differentiation circuits.
\newblock {\em Cell}, 139(3):512--522, 2009.

\bibitem{ouattara2010structure}
Djomangan~A Ouattara, Wassim Abou-Jaoud{\'e}, and Marcelle Kaufman.
\newblock From structure to dynamics: Frequency tuning in the p53-mdm2 network. ii: Differential and stochastic approaches.
\newblock {\em Journal of Theoretical Biology}, 264(4):1177--1189, 2010.

\bibitem{gonze2011molecular}
Didier Gonze, Wassim Abou-Jaoud{\'e}, Djomangan~Adama Ouattara, and Jos{\'e} Halloy.
\newblock How molecular should your molecular model be?: On the level of molecular detail required to simulate biological networks in systems and synthetic biology.
\newblock {\em Methods in Enzymology}, 487:171--215, 2011.

\bibitem{smadbeck2012stochastic}
Patrick Smadbeck and Yiannis Kaznessis.
\newblock Stochastic model reduction using a modified hill-type kinetic rate law.
\newblock {\em The Journal of Chemical Physics}, 137(23):234109, 2012.

\bibitem{riba2014combination}
Andrea Riba, Carla Bosia, Mariama El~Baroudi, Laura Ollino, and Michele Caselle.
\newblock A combination of transcriptional and microrna regulation improves the stability of the relative concentrations of target genes.
\newblock {\em PLOS Computational Biology}, 10(2):e1003490, 2014.

\bibitem{dovzhenok2015mathematical}
Andrey~A Dovzhenok, Mokryun Baek, Sookkyung Lim, and Christian~I Hong.
\newblock Mathematical modeling and validation of glucose compensation of the neurospora circadian clock.
\newblock {\em Biophysical Journal}, 108(7):1830--1839, 2015.

\bibitem{zhang2015negative}
Wei Zhang, Tianhai Tian, and Xiufen Zou.
\newblock Negative feedback contributes to the stochastic expression of the interferon-$\beta$ gene in virus-triggered type i interferon signaling pathways.
\newblock {\em Mathematical Biosciences}, 265:12--27, 2015.

\bibitem{schuh2020gene}
Lea Schuh, Michael Saint-Antoine, Eric~M Sanford, Benjamin~L Emert, Abhyudai Singh, Carsten Marr, Arjun Raj, and Yogesh Goyal.
\newblock Gene networks with transcriptional bursting recapitulate rare transient coordinated high expression states in cancer.
\newblock {\em Cell Systems}, 10(4):363--378, 2020.

\bibitem{jia2020small}
Chen Jia and Ramon Grima.
\newblock Small protein number effects in stochastic models of autoregulated bursty gene expression.
\newblock {\em The Journal of Chemical Physics}, 152(8):084115, 2020.

\bibitem{bersani2008deterministic}
Alberto~Maria Bersani, Enrico Bersani, and L~Mastroeni.
\newblock Deterministic and stochastic models of enzymatic networks—applications to pharmaceutical research.
\newblock {\em Computers \& Mathematics with Applications}, 55(5):879--888, 2008.

\bibitem{choi2017beyond}
Boseung Choi, Grzegorz~A Rempala, and Jae~Kyoung Kim.
\newblock Beyond the michaelis-menten equation: Accurate and efficient estimation of enzyme kinetic parameters.
\newblock {\em Scientific Reports}, 7(1):1--11, 2017.

\bibitem{d2017stability}
Matthew D’Alessandro, Stephen Beesley, Jae~Kyoung Kim, Zachary Jones, Rongmin Chen, Julie Wi, Kathleen Kyle, Daniel Vera, Michele Pagano, Richard Nowakowski, et~al.
\newblock Stability of wake-sleep cycles requires robust degradation of the period protein.
\newblock {\em Current Biology}, 27(22):3454--3467, 2017.

\bibitem{beesley2020wake}
Stephen Beesley, Dae~Wook Kim, Matthew D’Alessandro, Yuanhu Jin, Kwangjun Lee, Hyunjeong Joo, Yang Young, Robert~J Tomko, John Faulkner, Joshua Gamsby, et~al.
\newblock Wake-sleep cycles are severely disrupted by diseases affecting cytoplasmic homeostasis.
\newblock {\em Proceedings of the National Academy of Sciences of the United States of America}, 117(45):28402--28411, 2020.

\bibitem{bersani2020stochastic}
Alberto~Maria Bersani, Alessandro Borri, Francesco Carravetta, Gabriella Mavelli, and Pasquale Palumbo.
\newblock On a stochastic approach to model the double phosphorylation/dephosphorylation cycle.
\newblock {\em Mathematics and Mechanics of Complex Systems}, 8(4):261--285, 2020.

\bibitem{barik2008stochastic}
Debashis Barik, Mark~R Paul, William~T Baumann, Yang Cao, and John~J Tyson.
\newblock Stochastic simulation of enzyme-catalyzed reactions with disparate timescales.
\newblock {\em Biophysical Journal}, 95(8):3563--3574, 2008.

\bibitem{sanft2011legitimacy}
Kevin~R Sanft, Daniel~T Gillespie, and Linda~R Petzold.
\newblock Legitimacy of the stochastic michaelis--menten approximation.
\newblock {\em IET Systems Biology}, 5(1):58--69, 2011.

\bibitem{kang2019quasi}
Hye-Won Kang, Wasiur~R KhudaBukhsh, Heinz Koeppl, and Grzegorz~A Rempa{\l}a.
\newblock Quasi-steady-state approximations derived from the stochastic model of enzyme kinetics.
\newblock {\em Bulletin of Mathematical Biology}, 81(5):1303--1336, 2019.

\bibitem{kim2014validity}
Jae~Kyoung Kim, Kre{\v{s}}imir Josi{\'c}, and Matthew~R Bennett.
\newblock The validity of quasi-steady-state approximations in discrete stochastic simulations.
\newblock {\em Biophysical Journal}, 107(3):783--793, 2014.

\bibitem{kim2015relationship}
Jae~Kyoung Kim, Kre{\v{s}}imir Josi{\'c}, and Matthew~R Bennett.
\newblock The relationship between stochastic and deterministic quasi-steady state approximations.
\newblock {\em BMC Systems Biology}, 9(1):1--13, 2015.

\bibitem{BeyondMMPDE}
Seolah Shin, Seok~Joo Chae, Seunggyu Lee, and Jae~Kyoung Kim.
\newblock Beyond homogeneity: Assessing the validity of the michaelis–menten rate law in spatially heterogeneous environments.
\newblock {\em PLOS Computational Biology}, 20(6):1--22, 06 2024.

\bibitem{macnamara}
Shev MacNamara, Alberto~M. Bersani, Kevin Burrage, and Roger~B. Sidje.
\newblock Stochastic chemical kinetics and the total quasi-steady-state assumption: Application to the stochastic simulation algorithm and chemical master equation.
\newblock {\em The Journal of Chemical Physics}, 129(9):095105, 09 2008.

\bibitem{JITHINRAJ20141}
P.K. Jithinraj, Ushasi Roy, and Manoj Gopalakrishnan.
\newblock Zero-order ultrasensitivity: A study of criticality and fluctuations under the total quasi-steady state approximation in the linear noise regime.
\newblock {\em Journal of Theoretical Biology}, 344:1--11, 2014.

\bibitem{Herath}
Narmada Herath and Domitilla Del~Vecchio.
\newblock Reduced linear noise approximation for biochemical reaction networks with time-scale separation: The stochastic tqssa+.
\newblock {\em The Journal of Chemical Physics}, 148(9):094108, 03 2018.

\bibitem{kim2017reductionconserv}
Jae~Kyoung Kim, Grzegorz~A Rempala, and Hye-Won Kang.
\newblock Reduction for stochastic biochemical reaction networks with multiscale conservations.
\newblock {\em Multiscale Modeling and Simulation}, 15(4):1376--1403, 2017.

\bibitem{noethen2011tikhonov}
Lena Noethen and Sebastian Walcher.
\newblock Tikhonov’s theorem and quasi-steady state.
\newblock {\em Discrete and Continuous Dynamical Systems - B}, 16(3):945--961, 2011.

\bibitem{kim2017reduction}
Jae~Kyoung Kim and Eduardo~D Sontag.
\newblock Reduction of multiscale stochastic biochemical reaction networks using exact moment derivation.
\newblock {\em PLOS Computational Biology}, 13(6):e1005571, 2017.

\bibitem{song2021Universally}
Yun~Min Song, Hyukpyo Hong, and Jae~Kyoung Kim.
\newblock Universally valid reduction of multiscale stochastic biochemical systems using simple non-elementary propensities.
\newblock {\em PLOS Computational Biology}, 17(10):1--21, 10 2021.

\bibitem{Chae2025homohete}
Seok~Joo Chae, Seolah Shin, Kangmin Lee, Seunggyu Lee, and Jae~Kyoung Kim.
\newblock From homogeneity to heterogeneity: Refining stochastic simulations of gene regulation.
\newblock {\em Computational and Structural Biotechnology Journal}, 27:411--422, Jan 2025.

\bibitem{borghans1996extending}
Jos{\'e}~AM Borghans, Rob~J De~Boer, and Lee~A Segel.
\newblock Extending the quasi-steady state approximation by changing variables.
\newblock {\em Bulletin of Mathematical Biology}, 58(1):43--63, 1996.

\bibitem{schnell2002enzyme}
Santiago Schnell and Philip~K Maini.
\newblock Enzyme kinetics far from the standard quasi-steady-state and equilibrium approximations.
\newblock {\em Mathematical and Computer Modelling}, 35(1-2):137--144, 2002.

\bibitem{tzafriri2003michaelis}
A~Rami Tzafriri.
\newblock Michaelis-menten kinetics at high enzyme concentrations.
\newblock {\em Bulletin of Mathematical Biology}, 65(6):1111--1129, 2003.

\bibitem{frank2018quasi}
Martin Frank, Christian Lax, Sebastian Walcher, and Olaf Wittich.
\newblock Quasi-steady state reduction for the michaelis--menten reaction--diffusion system.
\newblock {\em Journal of Mathematical Chemistry}, 56:1759--1781, 2018.

\bibitem{kalachev2007reduction}
Leonid~V Kalachev, Hans~G Kaper, Tasso~J Kaper, Nikola Popovic, and Antonios Zagaris.
\newblock Reduction for michaelis-menten-henri kinetics in the presence of diffusion.
\newblock {\em Electronic Journal of Differential Equations (EJDE)[electronic only]}, 2007:155--184, 2007.

\bibitem{ETS}
Roktaek Lim, Thomas L.~P. Martin, Junghun Chae, Woo~Joong Kim, Cheol-Min Ghim, and Pan-Jun Kim.
\newblock Generalized michaelis–menten rate law with time-varying molecular concentrations.
\newblock {\em PLOS Computational Biology}, 19(12):1--18, 12 2023.

\bibitem{Chae_2025}
Junghun Chae, Roktaek Lim, Thomas~L.P. Martin, Cheol-Min Ghim, and Pan-Jun Kim.
\newblock Enlightening the blind spot of the michaelis–menten rate law: The role of relaxation dynamics in molecular complex formation.
\newblock {\em Journal of Theoretical Biology}, 597:111989, January 2025.

\end{thebibliography}






\end{document}